\documentclass[11pt]{article}
\usepackage{graphicx}
\usepackage{amsmath}
\usepackage{amssymb}
\usepackage{amsxtra}

\title{Statistical analyses of antimatter domains, created by nonhomogeneous baryosynthesis in a baryon asymmetrical Universe}
\author{Maxim Yu. Khlopov\\Institute of Physics, Southern Federal University,\\ Stachki 194 Rostov on Don 344090, Russia\\ Université de Paris, CNRS, Astroparticule et Cosmologie\\ F-75013 Paris, France, and \\National Research Nuclear University "MEPHI" 115409 Moscow, Russia\\
\\ email: khlopov@apc.in2p3.fr\\O.M. Lecian\\ Sapienza University of Rome, Faculty of Medicine and Pharmacy,\\ V.le Regina Elena, 324- 00185 Rome, Italy\\
Sapienza University of Rome, Faculty of Medicine and Dentistry,\\ P.zzale A. Moro, 5- 00185, Rome, Italy\\ email: orchideamaria.lecian@uniroma1.it}

\begin{document}
\maketitle

\begin{abstract}
 Within the framework of scenarios of nonhomogeneous baryosynthesis, the formation of macroscopic antimatter
domains is predicted in a matter-antimatter asymmetrical Universe.
The properties of antimatter within the domains are outlined; the matter-antimatter boundary interactions are studied.
The correlation functions for two astrophysical objects are calculated.
The theoretical expression in the limiting process of the two-points correlation function of an astrophysical object and an antibaryon is derived.
\end{abstract}



\section{Introduction}
The origin of the baryon asymmetry of the 
 Universe is explained in the now Standard
 cosmology by the mechanism of baryosynthesis. If baryon excess generation is nonhomogeneous, the appearance of domains with antibaryon excess can be predicted in baryon asymmetrical Universe.\\
In such non-trivial baryosynthesis frameworks, we study evolution of antimatter domains according to their dependence on the size and antimatter densities within them.\\
The boundary conditions for antimatter domains are determined through the interaction with the surrounding  baryonic medium.\\
Within the analysis, new classifications for antibaryon domains, which can evolve in antimatter globular clusters, are in order.\\
Differences must be discussed within the relativistic framework chosen, the nucleosynthesis processes, the description of the surrounding matter medium, the confrontation with the experimental data within the observational framework. The space-time-evolution of antimatter domains and the correlation functions are described within the nucleon-antinucleon boundary interactions.\\
The manuscript is organized as follows.\\
We consider formation of antibaryon domains in the spontaneous CP-symmetry-breaking scenario (Section \ref{section2}) and in the model of spontaneous baryosynthesis. Evolution of such domains is determined by nucleon-antinucleon interaction at the boundaries of antimatter domains.\\
 We deduce correlation functions for the celestial objects, predicted in these scenarios.\\
Them manuscript is organized as follows.\\
In Section \ref{section2}, the symmetry-breaking scenario have been recalled.\\
In Section \ref{section3}, the cosmological implications have been studied.\\
In Section \ref{section4}, the spontaneous baryosynthesis process has been analyzed.\\
I Section \ref{section4a}, different antimatter spacetime distributions have been presented.\\
In Section \ref{section5}, antimatter interactions have been studied.\\
Is Section \ref{section6}, nucleon-antinucleon interactions have been codified.\\
In Section \ref{section7}, correlation function for celestial objects have been analytically calculated.\\
Is Section \ref{section8}, brief outlook and perspectives have been outlined.\\
Concluding remarks end the paper.

\section{Symmetry-breaking scenario\label{section2}}
The symmetry-breaking scenarios have been stusied in \cite{ref1}-\cite{ref2c}.\\
The spontaneous CP violation is described \cite{ref1} after the Lagrangean potential density
\begin{equation}\label{eq1}
\begin{split}
V(\phi_1, \phi_2, \chi)&=-\mu_1^2(\phi_1^+\phi_1+\phi_2^+\phi_2)+\lambda_1[(\phi_1^+\phi_1)^2+(\phi_2^+\phi_2)^2]+2\lambda_3(\phi_1^+\\
&\phi_1)(\phi_2^+\phi_2)(\phi_1^+\phi_2)+2\lambda_4(\phi_1^+\phi_2)(\phi_2^+\phi_1)+\lambda_5[(\phi_1^+\phi_2)^2+h.c.]+\\
&\lambda_6(\phi_1^+\phi_1+\phi_2^+\phi_2)(\phi_1^+\phi_2+\phi_2^+\phi_1)-\mu_2^2\chi^+\chi+\delta(\chi^+\\
&\chi)^2+2\alpha(\chi^+\chi)(\phi_1^+\phi_1+\phi_2^+\phi_2)+2\beta[(\phi_1^+\chi)(\chi^+\phi_1)+(\phi_2^+\chi)(\chi^+\phi_2)].
\end{split}
\end{equation}
As a result, an effective low-energy electroweak $SU(2)\otimes U(1)$ theory is achieved, allowing for a
GUT spontaneous $CP$ violation.\\
The formation of vacuum structures separated from the rest of the matter universe by domain walls follows.

The size of the domains
 is calculated to grow with the evolution of the Universe.\\
The behavior is calculated not to affect the evolution of the Universe if the volume energy $\tilde{\rho(V)}$ density of the walls for\\
$\tilde{\rho(V)}\sim \sigma^2_\phi T^4/\tilde{h}$,\\
with $\tilde{h}$ value of the scalar coupling constant.\\


In Eq. (\ref{eq1}), for the three effective scalar fields

, the CP violation is achieved with complex vev's
and vacuum domain structures appear with opposite CP violation sign: walls are predicted to be massive, and the size of walls is predicted to grow \cite{ref1}.\\

A $CP$-invariant Lagrangian density can be assumed of the form \cite{ref2}
\begin{equation}
L=(\partial\phi)^2-\lambda^2(\phi^2-\chi^2)^2+\bar{\psi}(i\partial-m-ig\gamma_5\phi)\psi
\end{equation}
in which the vacuum is characterized by the values $<\phi>=\sigma\eta$, with $\sigma=\pm1$.\\
The rotation $\psi\rightarrow e^{i\alpha\gamma_5}\psi$, with $tg 2\alpha=-g\sigma\eta/m$\\
induces the appearance in the Lagrangian density of two terms with opposite CP symmetry. The sign of the phase depends on $\sigma$, with
 $\lambda\le 1$, for which 
$1\le \eta$.\\
For the Lagrangian density \cite{ref3}
\begin{equation}
L=(\partial\chi)^2-\frac{1}{2}m_\chi^2\chi^2-4\sigma\lambda2\chi\eta^3-\lambda^2\chi^4 +\bar{\psi}(i\hat{\partial}-M-i\frac{gm}{M}\gamma_5\chi-\frac{g^2\sigma\eta}{M}\chi)\psi.
\end{equation}
a $CP$ violation can be achieved after the substitution $\phi=\chi+\sigma\eta$.\\
For the Lagrangian potential
\begin{equation}
V(\chi)=-m^2_\chi\chi^*\chi+\lambda_\chi(\chi^*\chi)^2+V_0,
\end{equation}
with
$\chi=\frac{f}{\sqrt2}e^{\frac{i\alpha}{f}}$,
a $U(1)$ symmetry breaking is achieved, with 
 $\theta=\alpha/f$.\\
The domain wall problem can be solved after the Kuzmin-Shaposhnikov-Tkachev mechanism.
\section{Implications in cosmology\label{section3}}
Several phenomena can be looked for following the described mechanisms.\\
The research for antinuclei in cosmic rays is analyzed as a possible outcome of the model.\\
The research for annihilation products constitutes a further verification procedure for the theoretical framework.\\ 
In particular, annihilation at rest on Relativistic background is to be studied.\\
The
 annihilation of small-scale domains is a further investigation theme. It can be achieved within
the {\it{thin-boundary approximation}}.\\
Moreover, at different times, the {\it{diffusion of the baryon charge}} is determined after different processes.

\section{Spontaneous baryosynthesis\label{section4}}
A spontaneous baryosynthesis allowing for the possibility of sufficiently large domains through proper combination of effects of inflation and baryosynthesis is described after the choice of fields

$\chi\equiv\frac{f}{\sqrt{2}}e^\theta$,
for which the variance reads
\begin{equation}
<\delta\theta>=\frac{H^3t}{4\pi^2f^2}.
\end{equation}
as in \cite{a}, \cite{b}, \cite{ref6a}, \cite{ref7}, \cite{ref8}.
This way, the {\bf{probability for the existence of antimatter particles}} is set.\\
The number of objects $\tilde{N}(t)-\tilde{N}(t_0)$ is calculated as
\begin{equation}
\tilde{N}(t)-\tilde{N}(t_0)\equiv{\int_{t_0}}^{t_\iota}P(\chi)\ln\chi d\chi(t),
\end{equation}
with $P(\chi)$ including variance.\\
The evaluation of the number of antibaryons is performed after the use of the quantity
$\tilde{H}$, i.e. the Hubble-radius function, and after the definition of {\it{effective quantities}}
$\Delta f_{eff}$, i.e. the effective (time-dependent) phase function, and
\begin{equation}\label{feff}
f_{eff}=f\sqrt{1+\frac{g_{\phi\chi}M_{Pl}}{12\pi\lambda}(N_c-N)},
\end{equation}i.e. the effective phase, with
$N$ the e-foldings at inflation.\\


\normalsize
  
The following consequences are extracted.\\  
If the density is so low that nucleosynthesis is not possible, low density antimatter domains contain only antiprotons (and positrons).\\
High density antimatter domains contain antiprotons and antihelium.\\
Heavy elements can appear in stellar nucleosynthesis, or in the high-density antimatter domains.\\
Strong non-homogeneity in antibaryons might imply (probably as a necessary condition) strong non-homogeneity for baryons, and produce some exotic results in nucleosynthesis.
\section{Angular dispersion of the angular variable\label{section4a}}
For a PBH,
\begin{equation}
M_{min}=f(\frac{m_{Pl}}{\Lambda})^2
\end{equation}
for which
\begin{equation}
    m=\frac{\Lambda^2}{f}\ge \frac{m_{Pl}}{M_{min}}.
\end{equation}
The corresponding frequency dispersions are evaluated as
\begin{equation}
    \delta\omega=\frac{H_{infl}}{2\pi f}= \frac{H_{infl}m_{Pl}^2}{2\pi \Lambda^2M_{min}}\frac{\hbar}{c^2}\simeq 0.1\dot10^{-67}s^{-1}
\end{equation}
with an angular dispersion $\delta \theta/c\simeq\ge0.01\dot10^{-75}$ 
\subsection{Binomial spacetime antimatter distribution}
In the hypothesis antibaryons are described as following a binomial \cite{refc1} statistical spacetime distribution, the number of antibaryons
$\tilde{N}$ contained in an antimatter domain reads
\begin{gather}
\scalebox{0.75}{$
   \begin{aligned}
\tilde{N}(k)-\tilde{N}_0(k)\simeq & \sum_k \frac{1}{(k!)(1-k)!}\frac{\chi_{t_a}}{\chi_{t_0}}\frac{2}{\Delta f_{eff}(t; t_a, t_i, t_0)}\left(\frac{(-2)}{4\pi^2}\cdot
\left[ln\frac{L_u e^{H_c(t_c-t_0)}-e^{H_0t_0}}{l}\right]
\right)^k(t)^{k-3}
\end{aligned}$}
\end{gather}
which is described after the effective quantities $f_{eff}$ and those defined after the Hubble function $H$.

\subsection{Poisson space-time antimatter statistical distribution}


In the hypothesis antibaryons are described as following a Poisson statistical space-time distribution, the number of antibaryons
$\tilde{N}$ contained in an antimatter domain reads
\begin{gather}
\scalebox{0.75}{$
   \begin{aligned}
\tilde{N}(k)-\tilde{N}_0(k)(\Delta t)\simeq\sum_n\sum_k\frac{k^n e^k}{n!}\frac{\chi_{t_a}}{\chi_{t_0}}\frac{2}{\Delta f_{eff}(t; t_a, t_i, t_0)}\cdot\left(\frac{(-2)}{4\pi^2}\left[ln\frac{L_u e^{\tilde{H}_c(t_c-t_0)}-e^{\tilde{H}_0t_0}}{l}\right]\right)^k(t)^{k-3},
\end{aligned}$}
\end{gather}
 which is described after the effective quantities $f_{eff}$ and those defined after the Hubble function $H$. 
  
\subsection{Bernoulli spacetime antimatter distribution}
In the hypothesis antibaryons are described as following a Bernoulli statistical spacetime distribution, the number of antibaryons
$\tilde{N}$ contained in an antimatter domain reads
\begin{gather}
\scalebox{0.75}{$
   \begin{aligned}
\tilde{N}(k)-\tilde{N}_0(k)\simeq & \frac{1}{(k!)(1-k)!}\frac{\chi_{t_a}}{\chi_{t_0}}\frac{2}{\Delta f_{eff}(t; t_a, t_i, t_0)}\left(\frac{(-2)}{4\pi^2}\cdot\left[\ln\frac{L_u e^{H_c(t_c-t_0)}-e^{H_0t_0}}{l}\right]\right)^k(t)^{k-3}
\end{aligned}$}
\end{gather}
which is described after the effective quantities $f_{eff}$ and those defined after the Hubble function $H$.

\section{Antimatter domains and antibaryons interactions\label{section5}}
At the
{\it{radiation-dominated era}}

within the cosmological evolution, the dominant contribution to the total energy is due to photons.\\
In the case of 
\itshape{low density antimatter domains}\upshape ,
the contribution of the density of antibaryons $\rho_{\overline{B}}$ is smaller than the contribution due to the radiation $\rho_\gamma$ even at the matter-dominated stage.\\

In a FRW Universe, within its thermal history, for $T<100 keV$, only photons as a dominant components are considered.\\

At the
{\it{matter-dominated era and following}}

within a \itshape{non-homogeneous}\upshape $\ \ $scenario,
$\rho_{DM}>\rho_B$, with $\rho\equiv\rho(x)$.\\
The creation of high density antibaryon domains can be accompanied by similar increase in baryon density in the surrounding medium. Therefore outside high density antimatter domain baryonic density may be also higher than DM density
$\rho_{B}(x)>\rho_{DM}(x)$.\\
In the case of \itshape{low density antimatter domains:}\upshape \\
the total density is such that $\rho_{\overline{B}}+\rho_\gamma$, and
$\rho_{\overline{B}}<\rho_\gamma$, with
$\rho_{dm}>\rho_B$\\

\section{Nucleon-antinucleon interaction studies\label{section6}}
Within the framework of the studies of nucleons-antinucleons interaction, several schematizations are possible.\\ 
In the case of  {\it{proton-antiproton annihilation probability}}, the  {\it{limiting process}} and the {\it{theoretical formulation}} can be studied.\\
Let $P(\bar{p})$ be the probability of existence of one antiproton of mass $m_p$, with $m_p$ being the proton mass, in the spherical shell of section $r_I$, of (antimatter)-density $\rho_I$, delimiting the antimatter domain, in which the interaction takes place $P(\bar{p})\equiv 3Nm_p /(r_I\rho_I)$. This way, the\\
interaction probability reads\\
$\tilde{P}_i\equiv\tilde{P}_{\bar{p}\rightarrow(d.c._i)}$\\ i.e. it constitutes the probability of antiproton $\bar{p}$ interaction with a proton $p$ in a chosen $i$ annihilation channel $a.c.$, possibly also depending on the chemical potential.\\
Let $\Delta t$ be the time interval considered,
under the most general hypotheses (most stringent constraint), $\Delta t\pm \delta t$, $\Delta t \simeq t_U\simeq 4\cdot 10^{17} s$, with
$t_U$ age of the universe, $\delta t$ to be set according to the particular phenomena considered. This way
$\bar{P}_{\bar{p}, i}(t, \Delta t)$. i.e. the probability of antiproton interaction, i.e. antiproton-proton annihilation (density), reads

\begin{equation}\bar{P}_{\bar{p}, i}\simeq\frac{1}{\Delta t}P_{\bar{p}}\tilde{P}_i
\end{equation}
As a second study, the

 {\it{nucleon-antinucleon interaction (annihilation) probabilities}} are evaluated after the
antinucleus $\bar{M}$ interaction probability $\bar{P}_{\bar{M, j}}(t, \Delta t)$ through the annihilation channel(s) $k$ as
\begin{equation}
\bar{P}_{\bar{M}, k}(t, \Delta t)\simeq\frac{1}{\Delta t}P_{\bar{A}}\tilde{P}_{\bar{A, k}}.
\end{equation}In these examples, all the probabilities are normalized as $[t^{-1}]$.\\
  
  The studies of nucleons-antinucleons interactions are
to be further specified for
non-trivial Relativistic scenarios
\itshape{ such as perturbed FRW
with the thermal history of the Universe,
i.e., also, according to the Standard Cosmological Principle}\upshape.\\
The non-trivial Relativistic scenarios are schematized as
at large scales asymptotically isotropic and homogeneous.\\
Further specifications can be in order in the case of
 non-trivial nucleosynthesis,
possibilities of surrounding media,
antibaryon-baryon annihilation. In the latter case, the
most stringent constraint follows after $\bar{P}$ evaluated for present times
in the description of reducing density in the limiting process of a low-density antimatter domain.

\subsection{An example}
In the example of low-density antimatter domains, 
 non-interacting antiprotons are described,
 boundary interactions are taken onto account, and
interaction with surrounding medium can be considered.\\
In particular, low-density antimatter domains can be surrounded by low-density matter regions.

\section{Correlation functions\label{section7}}
{\it{Two-point correlation functions}} $\tilde{C}_2$ for two antimatter domains $\alpha_1$ and $\alpha_2$ of size $>10^3M_{\odot}$ each can be within the present framework analytically calculated.\\
More in detail, on (homogeneous, isotropic) Minkowski-flat background, and under the hypothesis antimatter densities $\rho\equiv\tilde{N}/V$ following a Poisson space-time statistical distribution.\\
The two-point correlation function $\tilde{C}_2$ is defined as
\begin{equation}
d\tilde{C}_2(\alpha_1, \alpha_2)\equiv \rho^2(1+\xi(\mid \vec{r}_{\alpha_1\alpha_2}\mid))dV_1dV_2
\end{equation}
where
\begin{equation}
 \xi(\mid \vec{r}_{\alpha_1\alpha_2}\mid)\equiv \mid \vec{r}_{\alpha_1\alpha_2}\mid
 \end{equation}
defines the estimator, and
$\vec{r}_{\alpha_1\alpha_2}$ the distance of the two antimatter domains.\\
Given two antimatter domains $\alpha_1$ and $\alpha_2$
of volume $V_{\alpha_l}\equiv\frac{4}{3}\pi r_l^3$,
separated of a distance $\mid \vec{r}_{\alpha_1\alpha_2}\mid$
{\it{the correlation function is analytically integrated as}}
\begin{equation}
\tilde{C}_2(\alpha_1, \alpha_2)=2\pi\tilde{n}(n,k; \Delta f_{eff}, \tilde{H}; \Delta t)\mid \vec{r}_{\alpha_1\alpha_2}\mid\left(\frac{1}{r_2}+\frac{1}{r_1}\right)\tilde{H_c}^{2k}t^{4k-4}
\end{equation}
evaluated at the present time $t$, with
$\tilde{H_c}$ the effective Hubble-radius function.
\subsection{An example: the two-point correlation functions for an antimatter domain and another object}
It is possible to consider the {\it{limiting example}} of the
correlation function between an antimatter domain $\alpha_1$ and an antibaryon $\alpha_3$. The {\it{
Davies-Peebles estimator}}
for the macroscopic objects described in terms of density distribution and temperature distribution reads
\begin{equation}
\xi_{l, l' }\equiv\frac{\tilde{N}_{bin}}{\tilde{N}}\frac{D_l(\mid \vec{r}\mid)}{D_{l'}(\mid \vec{r}\mid)}-1,
\end{equation}
with
$\tilde{N}$ number of antibaryons in a low-density antimatter domain,
where the antimatter is assumed to be distributed according to a Poisson space-time statistical distribution, and
$\tilde{N}_{bin}$ the number of antibaryons in a low-density antimatter domain
where the antimatter is distributed according to a binomial space-time statistical distribution,\\
the quantity $D_l(\mid \vec{r}\mid)$ indicates the number of pairs of low-density appropriate-mass antimatter domains
within the geodesics (coordinate) interval distance $\left[r-\frac{dr}{2}, r+\frac{dr}{2}\right]$, the quantity
$D_l'(\mid \vec{r}\mid)$ indicates the number of pairs of objects between an antimatter domain and\\
an (\itshape{Poisson-distributed}\upshape) antibaryon on the coordinate geodesics.\\ 
For the Davies-Peebles estimator
\begin{equation}
\xi_{l, l'}\equiv\frac{\tilde{n}_{bin}(n,k; \Delta f_{eff}, \tilde{H}; \Delta t)}{\tilde{n}(n,k; \Delta f_{eff}, \tilde{H}; \Delta t)}\frac{D_l(\mid \vec{r}\mid)}{D_{l}'(\mid \vec{r}\mid)}-1
\end{equation}
within the use of statistical estimators, the time dependence $\tilde{H_c}^{2k}t^{4k-4}$ is suppressed, and
the \itshape{time dependence}\upshape $\ \ $is expressed after the ratio $\frac{\tilde{n}_{bin}(n,k; \Delta f_{eff}, \tilde{H}; \Delta t)}{\tilde{n}(n,k; \Delta f_{eff}, \tilde{H}; \Delta t)}$, i.e. on the different statistical antimatter space-time distributions and on their dependence on the
 $\tilde{H}$ Hubble-radius function, and on the
 $\Delta f_{eff}$ effective (time-dependent) phase function.

\subsection{Hamilton estimator}
The Hamilton estimator\\
$\tilde{\xi}_{l, l'}$
takes into account the difference in distances among the Binomial distribution and the Poisson distribution.
  
\section{Outlook and perspectives\label{section8}}  
In the case antimatter domains are described to be separated in a small angular distance, the Rubin-Limber correlation functions \cite{ref4}, \cite{ref5} for small angles can be used.\\
An analysis of the metric requiring a time evaluation after the time of the surface of last scattering can be analyzed also for different metrics, as in \cite{ref6}

\section{Concluding remarks\label{section9}}
Prediction of macroscopic antimatter in baryon asymmetrical Universe is based on rather specific choice of parameters of baryosynthesis. To make antimatter domains sufficiently large to survive in the baryon matter surrounding a nontrivial combination of baryosynthesis and inflation are needed. It may look like we study a highly improbable and very exotic case. However, on the other hand, positive evidence for existence of macroscopic antimatter in our Galaxy, which may appear in the searches of cosmic antinuclei in AMS02 \cite{poulin} would strongly favor models, predicting antimatter domains in baryon asymmetric Universe, and would make possible to select the narrow classes of models of inflation and baryosynthesis, as well as to specify their parameters with high precision \cite{MK}. In view of this possibility we started to develop in the present work statistical analysis of possible space distribution of antimatter domains with the account for their evolution.\\

Confrontation of the predicted distribution of antimatter domains with the observational data would be important for multimessenger test of the models of nonhomogeneous baryosynthesis. The observable signatures of this distribution is the important direction of our future studies. In particular, the most probable forms of the evolved antimatter in our Galaxy should be clarified in this analysis.

It should be noted that the mechanisms of generation of antibaryon excess in baryon asymmetrical Universe may be accompanied by formation of domain walls at the border of antimatter domains. If these closed walls start to dominate, before they enter the horizon, the corresponding domains, surrounded by walls would become closed worlds, separating from our Universe. This open question is another challenge for our future analysis

 \section*{Acknowledgements}
The work by MK has been supported by the grant of the Russian Science Foundation (Project No-18-12-00213-P).   


\begin{thebibliography}{99}
\bibitem{ref1} V.A. Kuzmin, M.E. Shaposhnikov, I.I. Tkachev, Matter-antimatter domains in the Universe: a solution of the vacuum walls problem, Phys. Lett. {\bf B 105}, 1 (1981).
\bibitem{ref2} Ya.B. Zeldovich, L.B. Okun, L.Yu.  Kobzarev, Cosmological Consequences of a Spontaneous Breakdown of a Discrete Symmetry, Zh.  Eksp.  Teor.  Fiz.  {\bf40}, 1 (1974).
\bibitem{ref2a} V.M. Chechetkin, M.G. Sapozhnikov, M.Yu. Khlopov, Ya.B. Zeldovich, Astrophysical aspects of antiproton interaction with ${}^{4}$He (antimatter in the universe)  Phys. Lett. {\bf B 118}, 329 (1982).
\bibitem{ref2b} V.M. Chechetkin, M.G. Sapozhnikov, M.Yu. Khlopov, Antiproton interactions with light elements as a test of GUT cosmology, Riv. N. Cim. {\bf 5}, 1 (1982).
\bibitem{ref2c} V.A. Kuzmin, I.I. Tkachev, M.E. Shaposhnikov, Are There Domains of Antimatter in the Universe?, Zh. Eksp. Teor. Fiz. Lett. {\bf 33}, 557 (1981).
\bibitem{ref3}  M.Yu. Khlopov, S.G. Rubin, A.S. Sakharov, Possible origin of antimatter regions in the baryon dominated universe, Phys. Rev. {\bf D 62}, 083505 (2000)-
\bibitem{a} M.Yu. Khlopov, S.G. Rubin, A.S. Sakharov, Possible origin of antimatter regions in the baryon dominated universe, Phys. Rev. {\bf D 62}, 083505 (2000).
\bibitem{b} M.Yu. Khlopov, S.G. Rubin, A.S. Sakharov, Antimatter regions in the baryon dominated universe, 14th Rencontres de Blois on Matter- Anti-matter Asymmetry  [hep-ph/0210012].
\bibitem{ref6a} A.D. Dolgov et al., Baryogenesis During Reheating in Natural Inflation and Comments on Spontaneous Baryogenesis, Phys. Rev. {\bf D 56}, 6155 (1997).
\bibitem{ref7} A.G. Cohen and D.B. Kaplan, Thermodynamic generation of the baryon asymmetry, Phys. Lett. {\bf B 199} (1987) 251. 
\bibitem{ref8} A.G. Cohen and D.B. Kaplan, Spontaneous baryogenesis, Nucl. Phys. {\bf B 308} (1988)
913.
\bibitem{refc1} J. Bernoulli, Ars Conjectandi, Opus Posthumum. Accedit Tractatus de Seriebus infinitis, et Epistola Gallice scripta de ludo Pilae recticularis; Impensis Thurnisiorum, Fratrum, Basel (1713).
\bibitem{ref4} V.C. Rubin, Fluctuations in the space distribution of galaxies, Proc. Natl. Acad. Sci. USA {\bf 40}, 541 (1954).
\bibitem{ref5} D.N. Limber, The Analysis of Counts of the Extragalactic Nebulae in Terms of a Fluctuating Density Field. II., Astrophys. J. {\bf 119}, 655 (1954).
\bibitem{ref6} A. Pontzen, A. Challinor, Linearization of homogeneous, nearly-isotropic cosmological models, Class. Quant. Grav. {\bf 28}, 185007 (2011), Eq. (52).
\bibitem{DolgovAM}   A.D.  Dolgov: Matter and antimatter in the universe, Nucl. Phys. Proc. Suppl. {\bf 113}, 40 (2002).
\bibitem{Dolgov2} A. Dolgov, J. Silk: Baryon isocurvature fluctuations at small scales and baryonic dark matter, Phys. Rev. D {\bf 47}, 4244 (1993). 
\bibitem{Dolgov3} A.D. Dolgov, M. Kawasaki,  N. Kevlishvili: Inhomogeneous  baryogenesis,  cosmic  antimatter,  and  dark  matter, Nucl. Phys. B {\bf 807}, 229 (2009).
\bibitem{poulin} V. Poulin, P. Salati, I. Cholis, M. Kamionkowski, J. Silk: Where do the AMS-02 anti- helium events come from? Phys. Rev. D {\bf 99}, 023016 (2019).
\bibitem{MK}
M.Y. Khlopov.  What comes after the Standard Model? Prog. Part. Nucl. Phys. {\bf 116},  103824, (2021).
\end{thebibliography}
\end{document}